\documentstyle[12pt]{article}

\newcommand{\nc}{\newcommand}
\nc{\beq}{\begin{equation}}
\nc{\eeq}{\end{equation}}
\nc{\beqa}{\begin{eqnarray}}
\nc{\eeqa}{\end{eqnarray}}

\def\gsim{\mathrel{\rlap{\lower4pt\hbox{\hskip1pt$\sim$}}
    \raise1pt\hbox{$>$}}}       

\input epsf
\newwrite\ffile\global\newcount\figno \global\figno=1

\def\writedef#1{}
\def\figin{\epsfcheck\figin}\def\figins{\epsfcheck\figins}
\def\epsfcheck{\ifx\epsfbox\UnDeFiNeD
\message{(NO epsf.tex, FIGURES WILL BE IGNORED)}
\gdef\figin##1{\vskip2in}\gdef\figins##1{\hskip.5in}
\else\message{(FIGURES WILL BE INCLUDED)}%
\gdef\figin##1{##1}\gdef\figins##1{##1}\fi}
\def\figinsert{}
\def\ifig#1#2#3{\xdef#1{fig.~\the\figno}
\writedef{#1\leftbracket fig.\noexpand~\the\figno}%
\figinsert\figin{\centerline{#3}}\medskip\centerline{\vbox{\baselineskip12pt
\advance\hsize by -1truein\center\footnotesize{  Fig.~\the\figno.} #2}}
\bigskip\endinsert\global\advance\figno by1}
\def\endinsert{}

\begin{document}



\title{\large{\bf Minimum Length from Quantum Mechanics
and Classical General Relativity}}

\author{Xavier~Calmet\thanks{calmet@theory.caltech.edu},
Michael~Graesser\thanks{graesser@theory.caltech.edu} and
Stephen~D.H~Hsu\thanks{Permanent address: Institute of
Theoretical Science and Department of Physics, University of
Oregon, Eugene OR 97403.
Email: hsu@duende.uoregon.edu} \\
\\
California Institute of Technology, Pasadena, CA 91125\\
}



\maketitle

\begin{picture}(0,0)(0,0)
\put(650,355){\tt OITS-751 CALT 68-2496}
\end{picture}
\vspace{-24pt}

\begin{abstract}
We derive fundamental limits on measurements of position, arising
from quantum mechanics and classical general relativity. First, we
show that any primitive probe or target used in an experiment must
be larger than the Planck length, $l_P$. This suggests a
Planck-size {\it minimum ball} of uncertainty in any measurement.
Next, we study interferometers (such as LIGO) whose precision is
much finer than the size of any individual components and hence
are not obviously limited by the minimum ball. Nevertheless, we
deduce a fundamental limit on their accuracy of order $l_P$. Our
results imply a {\it device independent} limit on possible
position measurements.
\end{abstract}

{\vfill \hfill OITS 751  CALT 68-2496 }


\newpage


It is widely believed that a minimum length, of order the Planck
length $l_P$, results from combining quantum mechanics and
classical general relativity \cite{minlength,clock,foam}. That is, no
operational procedure (experiment) exists which can measure a
distance less than of order $l_P$. The key ingredients used to
reach this conclusion are the uncertainty principle from quantum
mechanics, and gravitational collapse (black hole formation) from
general relativity.

A dynamical condition for gravitational collapse is given by the Hoop
Conjecture (HC) \cite{hoop}, due to Kip Thorne: if an amount of energy
$E$ is confined at any instant to a ball of size $R$, where $R < E$,
then that region will eventually evolve into a black hole\footnote{We
use natural units where $\hbar, c$ and Newton's constant (or $l_P$)
are unity. We also neglect numerical factors of order one.}. Recent
results on black hole production in particle collisions \cite{bhp}
show strong support for the HC, even in the least favorable instance
where all of the energy $E$ is in the kinetic energy of two particles
moving past each other at the speed of light.

From the HC and the uncertainty principle, we immediately deduce the
existence of a {\it minimum ball} of size $l_P$. Consider a particle
of energy $E$ which is not already a black hole. Its size $r$ must
satisfy \beq r \gsim {\rm \bf max} \left[\, 1/E\, ,\,E \, \right]~~,
\eeq where $\lambda_C \sim 1/E$ is its Compton wavelength and $E$
arises from the hoop conjecture. Minimization with respect to $E$
results in $r$ of order unity in Planck units\footnote{
In this paper we focus specifically on spacelike minimum intervals. By
Lorentz covariance, we might expect that no spacetime interval 
$ds^2$ can be
measured to accuracy better than $l_P^2$, but this is a more subtle
issue \cite{cov}.}, or $r \sim l_P$.
If the particle {\it is} a
black hole, then its radius grows with mass: $r \sim E \sim 1/
\lambda_C$. This relationship suggests that an experiment designed (in
the absence of gravity) to measure a short distance $l << l_P$ will
(in the presence of gravity) only be sensitive to distances
$1/l$. This is the classical counterpart to T-duality in string theory
\cite{duality}.

It is possible that quantum gravitational corrections modify 
the relation between $E$ and $R$ in the HC \cite{oldlit}. However, 
if $E$ is much larger than the Planck mass, 
and $R$ much larger than $l_P$, we expect
semiclassical considerations to be reliable. (Indeed, in two particle
collisions with center of mass energy much larger than the Planck
mass the black holes produced are semiclassical.) This means that the
existence of a minimum ball of size much smaller than $l_P$ does {\it not}
depend on quantum gravity - the energy required to confine a particle
to a region of size much smaller than $l_P$ would produce a large, 
semiclassical black hole.

Before proceeding further, we give a concrete model of minimum
length that will be useful later. Let the position operator $\hat{x}$ have
discrete eigenvalues $\{ x_i \}$, with the separation between
eigenvalues either of order $l_P$ or
smaller.
(For regularly distributed eigenvalues with
a constant separation, this would be equivalent to a spatial lattice.)
We do not mean to imply that nature implements minimum length in this particular
fashion - most likely, the physical mechanism is more complicated, 
and may involve,
for example, spacetime foam or strings. However, our concrete 
formulation lends itself
to detailed analysis. We show below that this formulation
cannot be excluded by any gedanken experiment, which is strong evidence for the
existence of a minimum length.

Quantization of position does not by itself imply quantization of
momentum. Conversely, a continuous spectrum of momentum does not imply
a continuous spectrum of position. In a formulation of
quantum mechanics on a regular spatial lattice, with spacing $a$
and size $L$, the momentum operator has eigenvalues which are
spaced by $1/L$. In the infinite volume limit the momentum operator can have
continuous eigenvalues even if the spatial lattice spacing is kept
fixed. This means that the displacement operator \beq \label{disp}
\hat{x} (t) - \hat{x} (0) = \hat{p}(0) {t \over M} \eeq does not
necessarily have discrete eigenvalues (the right hand side of
(\ref{disp}) assumes free evolution; we use the Heisenberg picture
throughout). Since the time evolution operator is unitary the
eigenvalues of $\hat{x}(t)$ are the same as $\hat{x}(0)$. Importantly
though, the spectrum of $\hat{x}(0)$ (or $\hat{x}(t)$) is completely
unrelated to the spectrum of the $\hat{p}(0)$, even though they are
related by (\ref{disp})
\footnote{In general the discrete spectrum of an observable $A(0)$
is unrelated to that of $A(t)-A(0)$. For instance, consider a
Hamiltonian $H=\alpha \sigma_z$ and $A(0)=\sigma_x$. Then the
spectrum of $A(0)$ is $\pm1/2$, but the spectrum of
$\sigma_x(t)-\sigma_x(0)$ is $\pm  2 \sin (\alpha t/2)$.}.
Consequently, we stress that {\em a measurement of the
displacement is a measurement of the spectrum of $\hat{p}(0)$ (for
free evolution) and does not provide information on the spectrum
of $\hat{x}$.} A measurement of arbitrarily small displacement
(\ref{disp}) does not exclude our model of minimum length. To
exclude it, one would have to measure a position eigenvalue $x$
and a nearby eigenvalue $x'$, with $|x - x'| << l_P$.

Many minimum length arguments (involving, e.g., a microscope or
scattering experiment \cite{minlength}) are
obviated by the simple observation of the minimum ball. However,
the existence of a minimum ball does not by itself preclude the
localization of a macroscopic object to very high precision.
Hence, one might attempt to measure the spectrum of $\hat{x}(0)$
through a time of flight experiment in which wavepackets of
primitive probes are bounced off of well-localised macroscopic
objects. Disregarding gravitational effects, the discrete spectrum
of $\hat{x}(0)$ is in principle obtainable this way. But,
detecting the discreteness of $\hat{x}(0)$ requires wavelengths
comparable to the eigenvalue spacing.  For eigenvalue spacing
comparable or smaller than $l_P$, gravitational effects cannot be
ignored, because the process produces minimal balls (black holes)
of size $l_P$ or larger. This suggests a direct measurement of the
position spectrum to accuracy better than $l_P$ is not possible.
%
The
failure here is due to
the use of probes with very short wavelength.

A different
class of instrument - the interferometer - is capable of measuring
distances much smaller than the size of any of its sub-components
\cite{LIGO}. An interferometer can measure a distance \beq
\label{interferometer} \Delta x ~\sim~ { \lambda \over b \sqrt{N}}
~\sim~ {L \over \tau \sqrt{N} \nu}~~, \eeq where $\lambda = 1 /
\nu$ is the wavelength of light used, $L$ is the length of each
arm, $\tau$ the time duration of the measurement, and $N$
the number of photons. More precisely, $\Delta x$ is the change over
the duration of the measurement in the relative
path lengths of
the two arms of the interferometer.
$b = \tau / L$ is the number of bounces
over which the phase difference builds, so (\ref{interferometer})
can also be written as \beq \Delta \Phi ~=~ {b \Delta x \over
\lambda} ~\sim~ {1 \over \sqrt{N}}~~, \eeq which expresses
saturation of the quantum mechanical uncertainty relationship
between the phase and number operators of a coherent state.

From (\ref{interferometer}) it appears that $\Delta x$ can be made
arbitrarily small relative to $\lambda$ by, e.g., taking the
number of bounces to infinity\footnote{The parametrically improved
sensitivity of the interferometer compared to the naive
expectation $\Delta x \sim \lambda/ b$ is achieved by monitoring
the intensity of the recombined beams at the output port.}. Were
this the case, we would have an experiment that, while still using
a wavelength $\lambda$ much larger than $l_P$,  could measure a
distance less than $l_P$ along one direction, albeit at the cost
of making the measured object (e.g., a gravity wave) large in the
time direction. This would contradict the existence of a {\it
minimum interval}, though not a minimum ball in spacetime.
(Another limit which increases the accuracy of the interferometer
is to take the number of photons $N$ to infinity, but this is more
directly constrained by gravitational collapse. Either limit is
ultimately bounded by the argument discussed below.)

A constraint which prevents an arbitrarily accurate measurement of
$\Delta x$ by an interferometer arises due to the Standard Quantum
Limit (SQL) and gravitational collapse. The SQL \cite{SQL} is
derived from the uncertainty principle (we give the derivation
below; it is not specific to interferometers, although see
\cite{noise}) and requires that \beq \label{SQL} \Delta x \geq
\sqrt{ t \over 2 M }~~, \eeq where $t$ is the time over
which the measurement occurs and $M$ the mass of the object whose
position is measured. In order to push $\Delta x$ below $l_P$, we
take $b$ and $t$ to be large. But from (\ref{SQL}) this
requires that $M$ be large as well. In order to avoid
gravitational collapse, the size $R$ of our measuring device must
also grow such that $R > M$. However, by causality $R$ cannot
exceed $t$. Any component of the device a distance greater than
$t$ away cannot affect the measurement, hence we should not
consider it part of the device. These considerations can be
summarized in the inequalities \beq \label{CGR} t > R > M
~~.\eeq Combined with the SQL (\ref{SQL}), they require $\Delta x
> 1$ in Planck units, or \beq \label{DLP} \Delta x > l_P~. \eeq
(Again, we neglect factors of order one.)

Notice that the considerations leading to (\ref{SQL}), (\ref{CGR})
and (\ref{DLP}) were in no way specific to an interferometer, and
hence are {\it device independent}. We repeat: no device subject
to the SQL, gravity and causality can exclude the quantization
of position on distances less than the Planck length.

It is important to emphasize that we are deducing a minimum length
which is parametrically of order $l_P$, but may be larger or
smaller by a numerical factor.  This point is relevant to the
question of whether an experimenter might be able to transmit the
result of the measurement before the formation of a closed trapped
surface, which prevents the escape of any signal. If we decrease
the minimum length by a numerical factor, the inequality
(\ref{SQL}) requires $M >> R$, so we force the experimenter to
work from deep inside an apparatus which has far exceeded the
criteria for gravitational collapse (i.e., it is much denser than
a black hole of the same size $R$ as the apparatus). For such an
apparatus a horizon will already exist before the measurement
begins. The radius of the horizon, which is of order $M$, is very
large compared to $R$, so that no signal can escape.

We now give the derivation of the Standard Quantum Limit. Consider
the Heisenberg operators for position $\hat{x} (t)$ and momentum
$\hat{p} (t)$ and recall the standard inequality \beq \label{UNC}
(\Delta A)^2 (\Delta B)^2 \geq  ~-{1 \over 4} ( \langle [
\hat{A}, \hat{B} ] \rangle )^2 ~~.
\eeq Suppose that the
position of a {\it free} test mass is measured at time $t=0$
 and {\em again} at a later time.
The
position operator at a later time $t$ is \beq \label{P} \hat{x}
(t) = \hat{x} (0) ~+~ \hat{p}(0) \frac{t}{M}~~. \eeq
The commutator between the position operators at $t=0$ and $t$
is \beq [ \hat{x} (0), \hat{x} (t)] ~=~ i {t \over M}~~,
\eeq so using (\ref{UNC}) we have \beq \vert \Delta x (0) \vert
\vert \Delta x(t) \vert \geq \frac{t}{2M}~~.\eeq
We see that at least one of the uncertainties $\Delta x(0)$ or $\Delta x(t)$
must be larger than of order $\sqrt{t/M}$.
As a measurement of the discreteness of $\hat{x}(0)$
requires {\em two} position measurements,
it is limited by the greater of $\Delta x(0)$ or $\Delta x(t)$,
that is, by
$\sqrt{t/M}$.

The assumption of a free test mass in the SQL derivation deserves
further scrutiny. One might imagine that specially designed
interactions with the test mass during the time interval $(0 \, ,
\, t)$ might alter the bound by extracting some of the
momentum uncertainty.  However, we now argue that if the mass $M$
is that of the entire experimental apparatus (as restricted by
causality above), the SQL applies.

As a simple model for interactions between the test mass $m_1$ and
the rest of the apparatus, imagine a spring connecting it to
another mass $m_2$. If $m_2 >> m_1$ the spring damps out the
uncertainty in the position of $m_1$ due to the position
measurement at $t$. (The time evolution of $\hat{x}(t)$ would
involve the harmonic oscillator potential, not just the free
kinetic energy used to obtain (\ref{P}).) We could further imagine
that $m_2$ is connected to other masses $m_i >> m_2$, etc.
However, this construction terminates, due to causality, with any
masses which are further than $t$ away from $m_1$: they are not
part of the experiment and can be neglected. Let the total mass of
the system of masses and springs be $M \sim \sum m_i$. There is an
uncertainty in the center of mass coordinate $x_{\rm cm}$ of this
system due to the measurement performed on $m_1$ at time $t=0$.
Using causality, we can show that $\hat{x}_{\rm cm} (t)$ evolves
freely as in (\ref{P}) with $M$ given by the total mass: (a)
anything outside the causal radius $t$ cannot affect the
experiment, so we can simply remove it from our gedanken universe
without changing the results, and (b) the position of an isolated
apparatus in an empty universe must evolve freely according to
(\ref{P}). The uncertainty $\Delta x_{\rm cm}$ contributes to
$\Delta x$, which one can see by writing $\hat{x}(t) =
\hat{x}_{\rm cm} (t) + \hat{y}(t)$, with $[\hat{x}_{\rm cm}(t),
\hat{y}(t)] = 0$. We obtain a bound on $\Delta x$ which is
independent of the specifics of the interactions - we need only
use the total mass $M$ of all objects which can interact with
$m_1$ during the measurement.

The argument of the previous paragraph focuses on the
center-of-mass degree of freedom, but there are classes of
experiments - such as interferometers - that are only sensitive to
relative changes in position. For free particle motion the minimal
length bound detailed above also applies to the relative
coordinate. However, one might imagine interactions involving the
relative degree of freedom that could limit the growth of
uncertainty. Still, there are fundamental limits on
the ability of an external potential to extract the uncertainty in
momentum introduced by the initial measurement of $x(t)$.
Let the Hamiltonian for each arm be that of a simple
oscillator \beq H = \frac{1}{2} m \dot{x}^2 + \frac{1}{2} m \nu^2
(x - x_0)^2~~~. \eeq The width of the ground state wavefunction is
$\sigma \sim 1/ \sqrt{m \nu}$, which must be less than $l_P$. In
natural units, this requires $m
>> \nu^{-1}$. The HC then requires $L > m >> \nu^{-1}$, which
contradicts the causality requirement that $L < \nu^{-1}$ - i.e.,
that the size of the arm (spring plus masses) not exceed the
oscillation time-scale. If the causality requirement is violated,
the system no longer behaves like an oscillator with a single
displacement degree of freedom $x$.

There has been considerable discussion in the literature
of defeating the SQL using contractive states \cite{Yuen,Ozawa} or
other Quantum Non-Demolition (QND) techniques
\cite{noise,AY,obar,Braginsky:2001am,BC}. Contractive states allow for
uncertainties in position that do not grow in time as rapidly as
(\ref{SQL}).
Naively this may seem to allow for an accurate measurement of
the discreteness of the position operator, but
recall that
two measurements of position are needed to
do this.
Straightforward algebra using the properties of contractive
states (see, e.g., \cite{Yuen})
shows that the
uncertainties in
two subsequent measurements of position are still
bounded by (11). Alternatively, this follows
directly from the Heisenberg operator equations of
motion, or more intuitively, because
for a given level of desired uncertainty
for {\em both} measurements,
the time between
measurements cannot be
arbitrarily
long, since for these states the uncertainty in position
eventually begins to grow.  We emphasize
that as (11), not (5), was essential to our derivation,
the use of contractive states will not allow for a measurement
of the discreteness of position to scales
less than the Planck scale.

Similarly, we note that the QND proposals typically
amount to measurements of the displacement operator (\ref{disp}),
or of the time-integrated force on a test mass, which would appear
on the right hand side of (\ref{disp}) if we had not assumed free
evolution. In measuring the displacement operator (but {\it not}
the position operator $\hat{x}(t)$ itself at different times),
correlations between the initial and final states can be used to
cancel the dependence on the initial state of the test mass.
However, we have argued that such measurements do not probe the
discretization of the position operator, and hence cannot address
the question of minimum length.

\bigskip

\noindent The authors would like to thank C. Bauer, V. Braginsky,
Y.Chen, M. Perelstein, J. Preskill, K.S. Thorne, A. Weinstein and
M. Wise for discussions. C. Bauer contributed to this work at an
early stage. This research supported in part under DOE contracts
DE-FG06-85ER40224 and DE-FG03-92ER40701.


\bigskip


\end{document}